\documentclass[preprint2]{aastex}
\begin{document}
\title{RR Lyrae and Type II Cepheid Variables Adhere to a Common Distance Relation}
\author{Daniel J. Majaess}
\affil{Saint Mary's University, Halifax, Nova Scotia, Canada}
\affil{The Abbey Ridge Observatory, Stillwater Lake, Nova Scotia, Canada}
\email{dmajaess@ap.smu.ca}

\begin{abstract}
Preliminary evidence is presented reaffirming that SX Phe, RR Lyrae, and Type II Cepheid variables may be characterized by a common Wesenheit period-magnitude relation, to first order.  Reliable distance estimates to RR Lyrae variables and Type II Cepheids are ascertained from a single $VI$-based reddening-free relation derived recently from OGLE photometry of LMC Type II Cepheids.  Distances are computed to RR Lyrae ($d\simeq260$ pc), and variables of its class in the galaxies IC 1613, M33, Fornax dSph, LMC, SMC, and the globular clusters M3, M15, M54, $\omega$ Cen, NGC 6441, and M92.  The results are consistent with literature estimates, and in the particular cases of the SMC, M33, and IC 1613, the distances agree with that inferred from classical Cepheids to within the uncertainties: no corrections were applied to account for differences in metallicity.  Moreover, no significant correlation was observed between the distances computed to RR Lyrae variables in $\omega$ Cen and their metallicity, despite a considerable spread in abundance across the sample. In sum, concerns regarding a sizeable metallicity effect are allayed when employing $VI$-based reddening-free Cepheid and RR Lyrae relations. 
 \end{abstract}


\section{Introduction}
The study of Cepheids and RR Lyrae variables has provided rich insight into countless facets of our Universe.  The stars are employed: to establish distances to globular clusters, the Galactic center, and to galaxies exhibiting a diverse set of morphologies from dwarf, irregular, giant elliptical, to spiral in nature \citep{ud01,pi06,ma06,mat06,mat09,fe07,fe08,gr08,gi08,sc09,ma09,ma09c}; to clarify properties of the Milky Way's spiral structure, bulge, and warped disk \citep{ta70,op88,ef97,be06,ma09,ma09b}; to constrain cosmological models by aiding to establish $H_0$ \citep{fm96,fr01,ta02}; to characterize extinction where such variables exist in the Galaxy and beyond \citep{cc85,tu01,lc07,ko08,ma08,ma09b,ma09c}; to deduce the sun's displacement from the Galactic plane \citep{sh18,fe68,ma09}; and to probe the age, chemistry, and dynamics of stellar populations \citep{tu96b,lu98,an02,an02b,mo06}, etc. 

An additional bond beyond the aforementioned successes is shared between RR Lyrae variables and Type II Cepheids, namely that the stars obey a common distance and Wesenheit period-magnitude relation.  

\begin{figure}[!t]
\includegraphics[width=7cm]{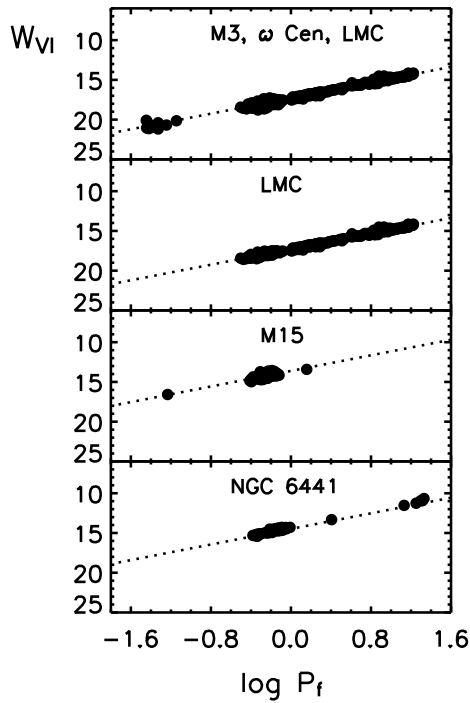}
\caption{\small{Wesenheit diagrams demonstrate that SX Phe, RR Lyrae, and Type II Cepheid variables follow a common period-magnitude relation.  Variable stars belonging to the globular clusters $\omega$ Cen and M3 were shifted in magnitude space to match the LMC.  The overplotted relation is Eqn.~\ref{eqn1} after adjusting the zero-point.  The fundamental mode period is plotted ($\log{P_f}$).}}
\label{fig1}
\end{figure}

\section{Analysis \& Discussion}
A Wesenheit period-magnitude diagram demonstrates the continuity from RRc to W Vir variables (Fig.~\ref{fig1}).  The Wesenheit function describing the OGLE LMC data is given by:
\begin{eqnarray}
\label{eqn1}
W_{VI}&=&V-\beta(V-I) \nonumber \\
W_{VI}&=&-2.45\log{P_f}+17.28 \nonumber \\
\end{eqnarray}
The relation is reddening-free and relatively insensitive to the width of the instability strip, hence the reduced scatter in Fig.~\ref{fig1}. Readers are referred to studies by \citet{vb68}, \citet{ma82}, \citet{op83}, \citet{mf09}, and \citet{tu09b} for an elaborate discussion on Wesenheit functions.  The colour coefficient used here, $\beta=2.55$, is that employed by \citet{fo07}.  RR Lyrae variables pulsating in the overtone were shifted by $\log{P_f}\simeq\log{P_o}+0.13$ so to yield the equivalent fundamental mode period \citep[e.g., see][]{so03,gr07}.  \citet{so08} convincingly demonstrated that the RV Tau subclass of Type II Cepheids do not follow a simple Wesenheit relation that also encompasses the BL Her and W Vir regimes \citep[see also][]{ma09}.  RV Tau variables were therefore excluded from the derived Wesenheit function which characterizes variables with pulsation periods $P \lesssim15^d$ (Eqn.~\ref{eqn1}).

\begin{figure}[!t]
\includegraphics[width=7.5cm]{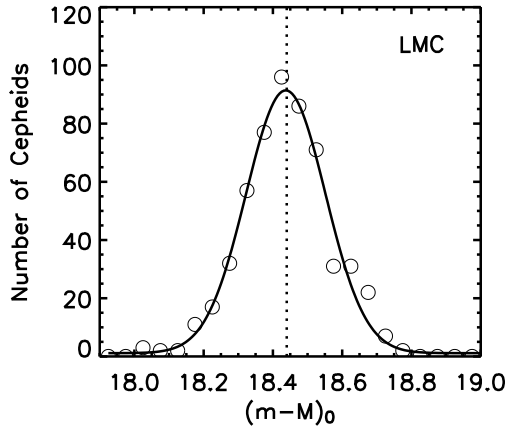}
\caption{\small{Applying the $VI$ reddening-free distance relation of \citet{ma08} to classical Cepheids observed in the LMC by \citet{se02} yields a distance modulus of $(m-M)_0=18.44\pm0.12$. }}
\label{fig5}
\end{figure}

A $VI$-based reddening-free Type II Cepheid relation \citep{ma09} was used to compute the distance to RR Lyrae variables in the galaxies IC 1613 \citep{do01}, M33 \citep{sa06}, Fornax dSph \citep{bw02,mg03}, LMC and SMC \citep{ud98,so02,so03,so09}, and the globular clusters M3 \citep{ben06,ha05}, M15 \citep{co08}, NGC 6441 \citep{lay99,pr03}, M54 \citep{ls00}, $\omega$ Cen \citep{we07}, and M92 \citep{ko01}. The resulting distances are summarized in Table~\ref{dgalaxies}, along with estimates from classical and Type II Cepheids, where possible.  The calibrators of the aforementioned relation were OGLE LMC Type II Cepheids \citep{ud99,so08}, with an adopted zero-point to the LMC established from classical Cepheids and other means \citep[$\sim18.50$,][]{gi00,fr01,be02,ma08}.  The distances to classical Cepheids were estimated using a Galactic calibration \citep{ma08} tied to a subsample of cluster Cepheids \citep[e.g.,][]{tu02} and new HST parallax measures \citep{be07}. Defining the relation strictly as a Galactic calibration is somewhat ambiguous given that Milky Way Cepheids appear to follow a galactocentric metallicity gradient \citep{an02,an02b}. The \citet{ma08} relation is tied to Galactic classical Cepheids that exhibit near solar abundances \citep{an02,an02b}.   Applying the \citet{ma08} relation to classical Cepheids observed in the LMC by \citet{se02} reaffirms the adopted zero-point ($(m-M)_0=18.44\pm0.12$, Fig.~\ref{fig5}).   No correction was applied to account for differences in metallicity between LMC and Galactic classical Cepheids owing to the present results and contested nature of the effect \citep[e.g.,][]{ud01,sa04,pi04,ma06,bo08,sc09,ma09c}.  A decrease of $\simeq0.08$ magnitudes would ensue if the correction proposed by \citet{sa04} or \citet{sc09} were adopted.

\begin{figure}[!t]
\includegraphics[width=7cm]{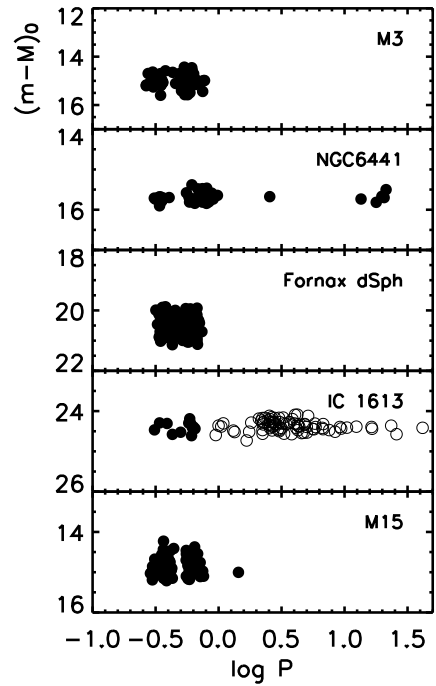}
\caption{\small{Period-distance diagrams for a subsample of objects studied.  Filled circles represent RR Lyrae variables and Type II Cepheids, while open circles denote classical Cepheids.}}
\label{fig3}
\end{figure}
\begin{figure}[!t]
\includegraphics[width=7.5cm]{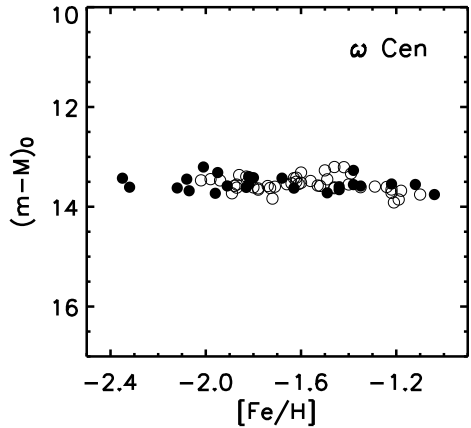}
\caption{\small{Abundance-distance diagram for RR Lyrae variables in the globular cluster $\omega$ Cen.  Open and filled circles are datapoints with metallicities inferred from $[Fe/H]_{hk}$ and $[Fe/H]_{\Delta S}$ methods \citep[see][]{re00}.  A formal fit yields a modest slope of $0.1\pm0.1$ mag dex$^{-1}$ and implies that metal poor RR Lyrae variables are brighter than metal poor ones.  The fit is also in agreement with no correlation.}}
\label{fig2}
\end{figure}

The distances cited in Table~\ref{dgalaxies} to the globular clusters are consistent with that found in the literature \citep[e.g.,][]{ha96}.  A subsample of period-distance diagrams demonstrate that the inferred distances are nearly constant across the entire period range examined (Fig.~\ref{fig3}). Moreover, distances computed to RR Lyrae variables in the SMC, M33, and IC 1613 agree with that inferred from classical Cepheids (Table~\ref{dgalaxies}). The results reaffirm that the slope and zero-point of $VI$ reddening-free relations are relatively insensitive to metallicity \citep[see also][]{ud01,pi04,ma08,ma09,ma09c}.  No corrections were made to account for differences in abundance.  Consider the result for the SMC, which is founded on copious numbers of catalogued RR Lyrae and classical Cepheid variables (OGLE).  SMC and Galactic classical Cepheids exhibit a sizeable metallicity difference ($\Delta [Fe/H]\simeq0.75$, \citet{mo06}).  By contrast, SMC RR Lyrae variables are analgous to or slightly more metal poor than their LMC counterparts \citep{ud00}.  If the metallicity corrections for RR Lyrae variables and classical Cepheids were equal yet opposite in sign as proposed in the literature (e.g., $\gamma_{-RR}\simeq+0.3$ \& $\gamma_{-Cep}\simeq-0.3$ mag dex$^{-1}$), then the distances computed for SMC RR Lyrae variables and classical Cepheids should display a considerable offset (say at least $\simeq0.25$ mag).  However, that is not supported by the evidence which instead implies a negligible offset separating the variable types ($0.02$ mag, Table~\ref{dgalaxies}). Similar conclusions are reached when analyzing distances to extragalactic RR Lyrae variables as inferred from a combined HST/HIP parallax for RR Lyrae \citep[see][]{ma09c}.  In sum, comparing Cepheids and RR Lyrae variables at a common zero-point offers a unique opportunity to constrain the effects of metallicity.  More work is needed here.  

RR Lyrae variables in $\omega$ Cen provide an additional test for the effects of metallicity on distance since the population exhibits a sizeable spread in metallicity at a common zero-point ($-1.0\ge [Fe/H] \ge-2.4$, \citet{re00}).  An abundance-distance diagram (Fig.~\ref{fig2}) compiled for RR Lyrae variables in $\omega$ Cen using $VI$ photometry from \citet{we07}, and abundance estimates from \citet{re00}, offers further evidence implying that $VI$-based reddening-free distance relations are relatively insensitive to metallicity.  A formal fit to data in Fig.~\ref{fig2} is in agreement with no dependence and yields a modest slope of $0.1\pm0.1$ mag dex$^{-1}$. If that slope is real metal poor RR Lyrae variables are brighter than metal rich ones.  Uncertainties linked to the cited slope could be mitigated by acquiring additional abundance estimates and obtaining $VI$ directly \citep[see][]{we07}.  A minor note is made that although the variable in $\omega$ Cen designated V164 is likely a Type II Cepheid (J2000 13:26:14.86 -47:21:15.17), the variable designated V109 may be anomalous or could belong to another variable class \citep[J2000 13:26:35.69 -47:32:47.03, see numbering in][]{we07}.

\begin{figure}[!t]
\includegraphics[width=7.5cm]{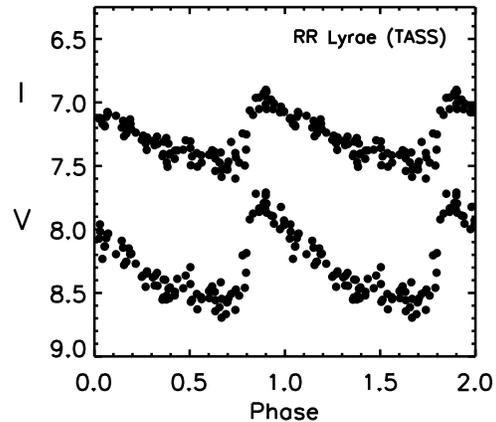}
\caption{\small{$V$ \& $I$ light-curves for RR Lyrae phased using equation~\ref{eqn2}.  The $V$-band photometry is offset by $+0.5$ mag.  The distance to RR Lyrae using \textit{The Amateur Sky Survey} photometry is $d\simeq260$ pc (see text).}}
\label{fig4}
\end{figure}

\begin{deluxetable}{lcccc}
\tabletypesize{\scriptsize}
\tablecaption{Distances to the sample.\label{dgalaxies}} 
\tablewidth{0pt}
\tablehead{\colhead{Object} &\colhead{$(m-M)_0$ (RR)} &\colhead{$(m-M)_0$ (TII)\tablenotemark{b}} &\colhead{$(m-M)_0$ (TI)} &\colhead{Photometry}}
\startdata
IC 1613 &$24.40\pm0.12$ &\nodata & $24.50\pm0.12$\tablenotemark{a} & (6) \\
 & \nodata & $24.52\pm0.16$ ($n=2$) & $24.35\pm0.09$ & (9) \\
SMC &$18.91\pm0.08$ & $18.85\pm0.11$ & $18.93\pm0.10$ & (7,8) \\
M33 &$24.54\pm0.14$ & \nodata & \nodata& (9) \\
& \nodata & $24.54$ ($n=1$) & $24.43\pm0.14$ (i) / $24.67\pm0.07$ (o)\tablenotemark{c}  & (10) \\
& \nodata & $24.5\pm0.3$ & $24.40\pm0.17$ (i) & (11) \\
Fornax dSph &$20.53\pm0.08$ & \nodata & \nodata & (5) \\
 &$20.57\pm0.11$ & \nodata & \nodata & (4) \\
M54 &$17.15\pm0.16$ & $17.13\pm0.06$ & \nodata & (3) \\
M92 &$14.57\pm0.12$ & \nodata  & \nodata & (2) \\
NGC 6441 & $15.70\pm0.06$ & $15.69\pm0.10$ &\nodata & (1) \\
& $15.74\pm0.11$ & \nodata &\nodata & (16) \\
M3 & $15.04\pm0.11$ & \nodata &\nodata & (12) \\
& $15.04\pm0.07$ & \nodata &\nodata & (15) \\
$\omega$ Cen & $13.58\pm0.14$ & \nodata &\nodata & (13) \\
M15 & $14.90\pm0.11$ & \nodata &\nodata & (14) \\
\enddata
\tablecomments{References: (1) \citet{pr03}, (2) \citet{ko01}, (3) \citet{ls00}, (4) \citet{bw02}, (5) \citet{mg03}, (6) \citet{do01}, (7) \citet{ud98}, (8) \citet{so02}, (9) \citet{sa06}, (10) \citet{sc09}, (11) \citet{ma01}, (12) \citet{ha05}, (13) \citet{we07}, (14) \citet{co08}, (15) \citet{ben06}, (16) \citet{lay99}.}
\tablenotetext{a,b,c}{See discussion \citet{do01}.  \textbf{***} Presently, the distance estimates for extragalactic Type II Cepheids should be interpreted cautiously owing to poor statistics and other concerns \citep{ma09c}. Distances to classical Cepheids occupying the outer (o) and inner (i) regions of M33.}
\end{deluxetable}

Equation 2 of \citet{ma09} was also employed to compute the distance to the brightest member of the variable class, RR Lyrae.  $VI$ photometry from \textit{\textit{The Amateur Sky Survey}} was utilized \citep{dr06}, although concerns persist regarding the survey's zero-point and the star's modulating amplitude.  Nevertheless, the resulting distance of $d\simeq260$ pc is consistent with the star's parallax as obtained using HST \citep[$d=262\pm14$ pc,][]{be02}, and within the uncertainties of the HIP value \citep{vl07,fe08}.  That reaffirms the robustness of the aforementioned relation to compute distances to variables of the RR Lyrae and Type II Cepheid class.  RR Lyrae's phased $V$ \& $I$ light-curves are displayed in Figure~\ref{fig4}.  An ephemeris from the GEOS RR Lyr database was adopted to phase the data \citep{bo02,lb04,lb07}, namely:
\begin{equation}
\label{eqn2}
JD_{max}=2442923.4193+0.5668378 \times E 
\end{equation}

The slope of the Wesenheit function derived from a combined sample of SX Phe, RR Lyrae, and Type II Cepheid variables detected in M3, $\omega$ Cen, and M15, is consistent with that determined from LMC RR Lyrae variables and Type II Cepheids (Fig.~\ref{fig1}).  Presently, the distances computed to SX Phe variables discovered in M3 and $\omega$ Cen via the $VI$ reddening-free Type II relation of \citet{ma09} are systemically offset.  However, the new Wesenheit relation performs better (Eqn.~\ref{eqn1}).  A reanalysis is anticipated once a sizeable sample of SX Phe variables become available.  Yet meanwhile Eqn.~\ref{eqn1} may be employed to evaluate the distances to SX Phe, RR Lyrae, BL Her, and W Vir variables. Uncertainties are expected to be on the order of 5-15\%.  Indeed, the correction factor ($\phi$) established by \citet{ma09} could be applied to Eqn.~\ref{eqn1} to permit the determination of distances to the RV Tau subclass of Type II Cepheids.  

Admittedly, further work is needed but the results are encouraging.

\section{Summary \& Future Work}
A single $VI$-based reddening-free relation may be employed to simultaneously provide reliable distances to RR Lyrae variables and Type II Cepheids. The  relation's viability is confirmed by demonstrating that distances to RR Lyrae variables in the globular clusters M3, M15, M54, $\omega$ Cen, M92, NGC 6441, and galaxies IC 1613, M33, Fornax dSph, LMC, and SMC, agree with values in the literature and from other means \citep[Table~\ref{dgalaxies}, see also][]{ha96}.  A distance was computed for the nearby star RR Lyrae ($d\simeq260$ pc) using mean $VI$ photometry provided by \textit{The Amateur Sky Survey}.  The estimate is consistent with the HST parallax for the star \citep[$d=262\pm14$ pc,][]{be02}.  The slope and zero-point of the $VI$-based relation appear relatively unaffected by metallicity to within the uncertainties (Table~\ref{dgalaxies} \& Fig.~\ref{fig1},~\ref{fig3}).  That assertion is supported by noting that although RR Lyrae variables in $\omega$ Cen exhibit a sizeable spread in metallicity ($-1.0\ge [Fe/H] \ge-2.4$, \citet{re00}), no statistically significant effect was observed on the computed distances (Fig.~\ref{fig2}).  Furthermore, the distances computed to RR Lyrae variables and classical Cepheids in the SMC, M33, and IC 1613 are consistent to within the uncertainties. No metallicity correction was applied.  Finally, SX Phe, RR Lyrae, and Type II Cepheids essentially follow a common Wesenheit period-magnitude relation, although poor statistics for the SX Phe variables currently limits an elaborate analysis (Fig.~\ref{fig1}).   

\begin{figure*}[!t]
\begin{center}
\includegraphics[width=14.5cm]{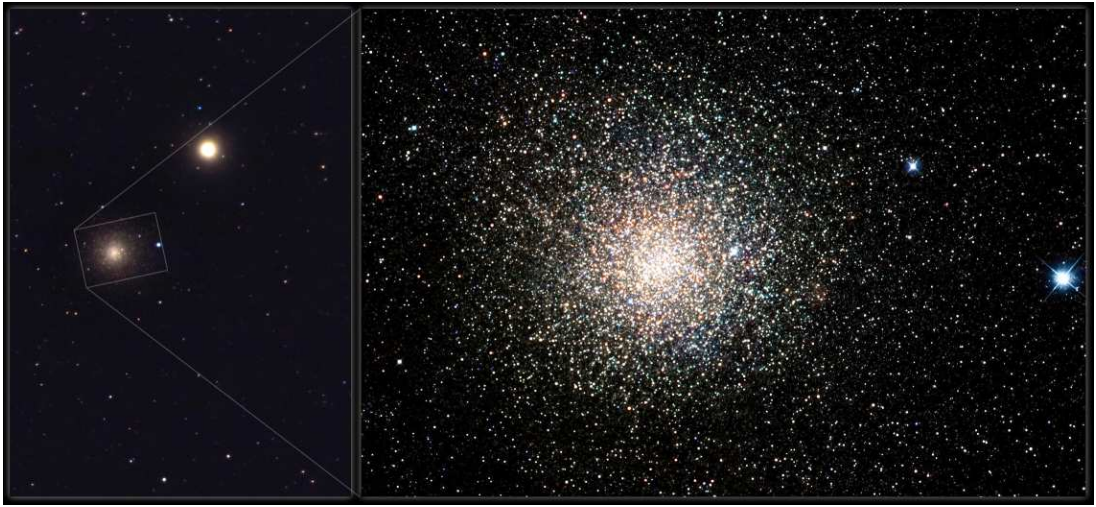}
\end{center}
\caption{\small{The metal rich globular cluster NGC 6441 observed through a 10\arcsec telescope (left) and the Hubble Space Telescope (right).  Image by Noel Carboni.}}
\label{fig6}
\end{figure*}

There remain numerous challenges and concerns to be addressed regarding the use of the distance indicators beyond the contested effects of metallicity \citep[e.g.,][]{ud01,sa04,pi04,ma06,bo08,sc09,ma09c}.  For example, achieving a common photometric standardization is difficult and systemic offsets may be introduced, particularly across a range in color \citep[e.g.,][]{tu90,sah06}. Yet another challenge is to establish a consensus on the effects of photometric contamination (e.g., blending, crowding) on the distances to variable stars in distant galaxies \citep{su99,fr01,mac01,mo00,mo01,mo02,ma09c}.  Increasing the presently small number of galaxies with Cepheids observed in both the central and less-crowded outer regions is therefore desirable \citep[e.g.,][]{ma06,sc09}.  Unfortunately, a degeneracy complicates matters since the effects of metallicity and crowding may act in the same sense and be of comparable magnitude.  Indeed, $R$ (the ratio of total to selective extinction) may also vary as a function of radial distance from the centers of galaxies in tandem with the metallicity gradient.  Efforts to disentangle the degeneracies are the subject of a study in preparation.  Further research is warranted to examine the implications of anomalous values of $R$ on the distances obtained from the standard candles \citep[e.g.,][]{ma01b,ud03}.  

Lastly, the continued discovery of extragalactic SX Phoenicis, RR Lyrae, and Cepheids at a common zero-point shall bolster our understanding and enable firm constraints to be placed on the metallicity effect \citep{sz06,po06,ma09c}.  So too will obtaining mean multiband photometry, particularly $VI$, for such variables in the field and globular clusters \citep{sa39,dw77,cl01,pr03,es04,es05c,es05b,es05,es09,ho05,mat06,ran07,rab07,co08}. A forthcoming study shall describe how related efforts are to be pursued from the Abbey-Ridge Observatory (ARO) \citep{la07,ma08b,tu09c}.  AAVSO members drawn toward similar research may be interested in a fellow member's study entitled: \textit{``Using a Small Telescope to Detect Variable Stars in Globular Cluster NGC 6779"} \citep{ho05}.  Modest telescopes may serve a pertinent role in variable star research \citep{pe80,pe86,sz03,pa06,tu05,tu09c}.   

\subsection*{acknowledgements}
\scriptsize{I am grateful to D. Weldrake, D. Bersier, G. Kopacki, V. Scowcroft, L. Macri, B. Pritzl, K. Sebo, A. Mackey, T. Corwin, A. Dolphin, A. Sarajedini, J. Hartman, J. Benk{\H o}, A. Layden, A. Udalski \& I. Soszy{\'n}ski (OGLE), whose comprehensive surveys were the foundation of the research, to the AAVSO and M. Saladyga, les individus au Centre de Donn\'ees astronomiques de Strasbourg et NASA ADS, L. Berdnikov, L. Szabados, J. F. Le Borgne, W. Renz, N. Carboni, and the RASC.  The following works facilitated the preparation of this study: \citet{wc84}, \citet{fm96}, \citet{fe99,fe01,fe08b}, \citet{fe69,fe76,fe02}, \citet{ho02}, \citet{wa02}, \citet{sz06,sz06b}, \citet{sm04}, and \cite{mar09}.}

\end{document}